\documentclass[a4paper]{jpconf} 
\usepackage{graphicx}
\begin{document}
\title{Fast imaging of single photons in quantum assisted optical interferometers}

\author[a]{Andrei Nomerotski$^{a}$, Jonathan Schiff$^{~a}$, Paul Stankus$^{a}$, Michael Keach$^{a}$, Alexander Parsells$^{a}$, Olli Saira$^{a}$, An\v{z}e Slosar$^{a}$, Stephen Vintskevich$^{b}$}
\address{[a] Brookhaven National Laboratory, Upton NY 11973, USA}
\address{[b] Moscow Institute of Physics and Technology,
Dolgoprudny, Moscow Region 141700, Russia}

\ead{anomerotski@bnl.gov}

\begin{abstract}
We describe a new technique of quantum astrometry, which potentially can improve the resolution of optical interferometers by orders of magnitude. The approach requires fast imaging of single photons with sub-nanosecond resolution, greatly benefiting from recent advances in photodetection technologies. We also describe results of first proof of principle measurements and lay out future plans.
\end{abstract}

\section{Introduction}

Improving angular resolution in astronomical instrumentation has long been pursued in order to advance astrometric precision.  Although  traditional optical interferometry is a very successful astronomical technique, the required optical path between different detection stations to interfere photons limits  baselines to hundreds of meters \cite{tenBrummelaar2005, Pedretti2009}.
This paper discusses the instrument requirements for a novel type of optical interferometer first proposed in our previous work \cite{stankus2020, Nomerotski2020_1} that utilizes quantum mechanical interference effects between two photons from two astronomical sources.
By using two sky sources, the proposed interferometer bypasses the traditional necessity of establishing a live optical path connecting detection sites, so the baseline distance can be made arbitrarily large, and consequently, an improvement of several orders of magnitude in angular resolution is in principle attainable.  This development can be considered as a practical variation of pioneering ideas described in the work of Gottesman, Jennewein and Croke in 2012 \cite{Gottesman2012} to employ a source of single photons to measure the photon phase difference between the receiving stations, which were further developed in the following references \cite{harvard1, harvard2}. At the same time, our ideas overlap with the Hanbury Brown \& Twiss~(HBT) intensity correlation astronomical technique~\cite{hbt, Foellmi2009}, which aims to resolve angular star dimensions employing two-photon enhancement effects.

There are many scientific opportunities that would benefit from substantial improvements in astrometric precision.  Just to list a few: testing  theories  of  gravity  by  direct  imaging  of  black  hole  accretion  discs, precision parallax and the cosmic distance ladder, mapping microlensing events, peculiar motions and dark matter; see \cite{stankus2020} for a more comprehensive discussion.  In this paper we briefly discuss the general setup of the interferometer in Section 2 and requirements for single photon detectors in Section 3. We then focus on the recent experimental results in Section 4.

%
\section{Two-photon interferometry}

\begin{figure}[h]
\begin{center}
\includegraphics[width=0.44\linewidth]{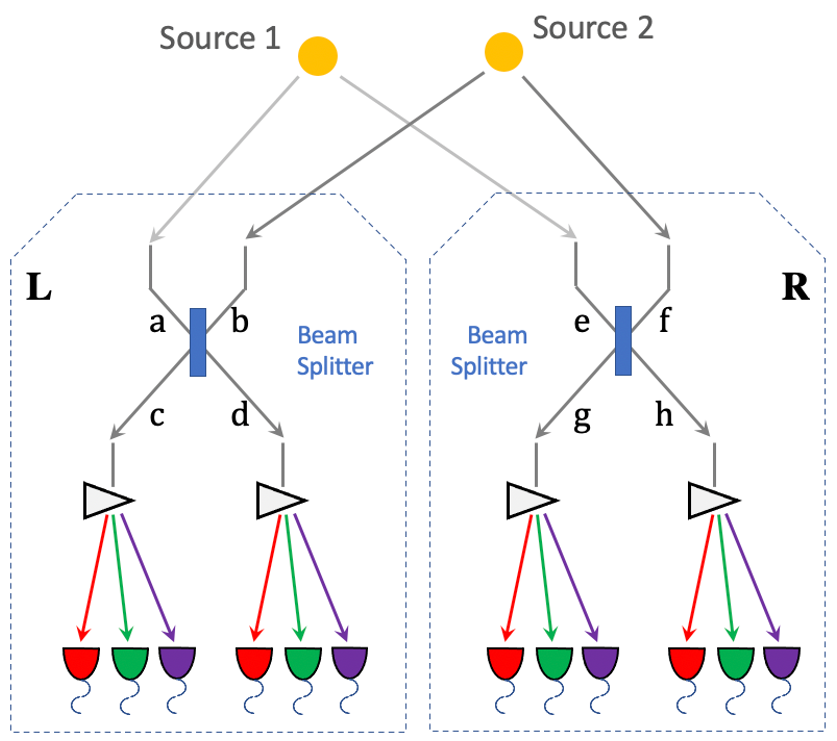}
\includegraphics[width=0.55\linewidth]{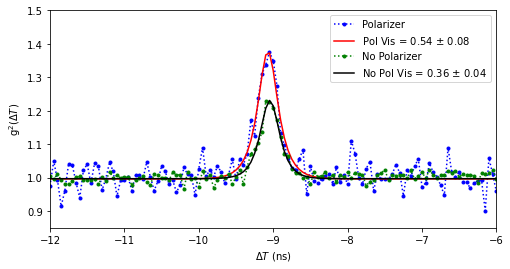}
\caption{Left: Two-photon amplitude interferometer. Right: Comparison of visibility in HBT peak with and without the source being polarized.}
\label{fig:idea}
\end{center}
\end{figure}

The basic arrangement of the novel interferometer is shown in the left panel of Figure~\ref{fig:idea}: the two sources~1 and~2 are both observed from each of two stations, \textbf{L} and~\textbf{R}.  Different optical schemes, using two or four telescopes, can be devised, but the key requirement is that photons from Source~1 be coupled into single spatial modes~$a$ at station~\textbf{L} and~$e$ at station~\textbf{R}; while those from Source~2 are separately coupled into the two single spatial modes $b$ and $f$ as shown.

The photon modes~$a$ and~$b$ at station~\textbf{L} are then brought to the inputs of a symmetric beam splitter, with output modes labelled $c$ and $d$; and the same for input modes $e$ and $f$ split onto output modes $g$ and $h$ at station~\textbf{R}.  The four outputs are then each viewed by a fast, single-photon sensitive detector.  We imagine that the light in each output port is spectrographically divided into small bins and each spectral bin then constitutes a separate experiment with four detectors.  
If the two photons are close enough together in both time and frequency, then due to quantum mechanical interference the pattern of coincidences between measurements at ``c'' and ``d'' in \textbf{L} and ``g'' and ``h'' in \textbf{R} will be sensitive to the {\em difference} in phase differences $(\delta_{1} - \delta_{2})$; and this in turn will be sensitive to the relative opening angle between the two sources.

In our previous work \cite{stankus2020} we derived that the corresponding coincidence probabilities are equal to:
\begin{eqnarray}
P(cg) & = & P(dh)  =   
       (1/8) (1+\cos(\delta_{1} - \delta_{2})) \nonumber \\
P(ch) & = & P(dg)  =   
       (1/8) (1-\cos(\delta_{1} - \delta_{2})).
\label{eq:outcomes_source_12}
\end{eqnarray}

%
\section{Requirements and options for photon detectors}

An important consideration is that the photons must be close enough in time {\em and} in frequency to efficiently interfere; or, formulating it differently, to be indistinguishable within the uncertainty principle $ \Delta t \cdot \Delta E \sim \hbar $. Converting energy to wavelength, the above expression is satisfied for $\Delta t \cdot \Delta \lambda = 10~\mathrm{ps} \cdot 0.2~\mathrm{nm}$ at 800~nm wavelength, setting useful target goals for the temporal and spectral resolutions.
Another important parameter for the imaging system is the photon detection efficiency, which needs to be as high as possible, since the two-photon coincidences used in the proposed approach have a quadratic dependence on it.

An efficient scheme of spectroscopic binning can be implemented by employing a traditional diffraction grating spectrometer where incoming photons are passed though a slit, dispersed, and then focused onto a linear array of single-photon detectors.
In our previous work we implemented spectroscopic binning employing a fast pixel camera \cite{Nomerotski2019, ASI} and determined the temporal and spectroscopic resolutions for pairs of registered photons in the context of quantum target detection \cite{Zhang2020, Svihra2020}.  We achieved $\Delta t \cdot \Delta \lambda = 5~\mathrm{ns} \cdot 0.5~\mathrm{nm}$ at 800 nm representing a useful benchmark for current technology. However, it is still three orders of magnitude from the resolution that is needed for the maximum visibility. Improvement of timing resolution appears to be the most straightforward way to achieve the targeted performance.

Several fast technologies, such as superconducting nanowire single photon detectors (SNSPD), single photon avalanche devices (SPAD) and streaking tubes, can be considered for this application \cite{Nomerotski2020_1}. In particular, SNSPD is an emerging quantum sensor technology which employs narrow, serpentine superconducting wires to detect single photons. A photon depositing its energy in the vicinity of the wire disrupts the superconductivity locally, inducing a voltage signal in the detection circuit  \cite{Divochiy2008, Zhu2020, Korzh2020}. The superconducting nanowire detectors have excellent photon detection efficiency, in excess of 90\%. Recent improvements in the detector geometry and readout circuitry have yielded the demonstrations of 3~ps timing resolution~\cite{Korzh2020} and small-scale pixel arrays \cite{Miyajima2018,Allmaras2020}. In the next section we discuss the first experimental results obtained with this technology.

Another promising technology, SPAD sensors are based on silicon diodes with engineered junction breakdown. They produce fast pulses of big enough amplitude to operate with single-photon sensitivity \cite{Gasparini2017, Perenzoni2016, Lee2018}. These devices have excellent timing resolution, which can be as good as 10~ps for single-channel devices, and most importantly, good potential for scalability with multi-channel imagers already reported \cite{Gasparini2017, Morimoto2020}.

%
\section{Experimental setup and first results}

We performed an initial verification of ideas described above with a simplified scheme employing two quasi-thermal sources, which were fiber-coupled to 1-to-2 beam splitters.  Light from the sources was then mixed in two 2-to-2 beam splitters and was directed to four single photon detectors.  All beam splitters in the experiments were coupled to single mode fibers, thus implementing a fiber-coupled version of the layout shown in Figure \ref{fig:idea}. Two low pressure argon lamps were used as narrow-band sources of thermal photons by selecting a narrow spectral line at $794.818$~nm with a $794.9 \pm 1$~nm filter. The source coherence time was evaluated by observing the Hanbury Brown and Twiss (HBT) effect of intensity correlations. The right panel of Figure~\ref{fig:idea} gives an example of two HBT peaks, with and without a linear polarizer at the source, fit with a Lorentzian function convoluted with experimental time resolution. As expected the visibility for the case with polarizer in place is approximately twice as large as without it.  The result in Figure~\ref{fig:idea} represents coincidences observed between photons at a specific pair of detectors but similar peaks were observed for all six combinations of two detectors.

\begin{figure}[h]
\begin{center}
\includegraphics[width=0.48\linewidth]{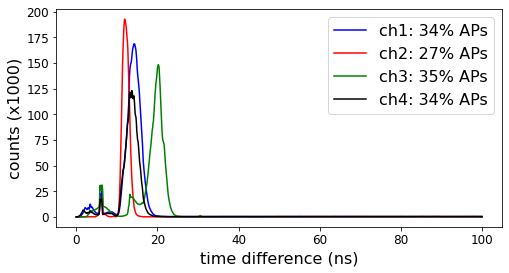}
\includegraphics[width=0.51\linewidth]{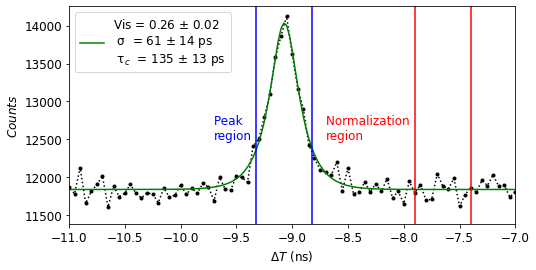}
\caption{Left: Distribution of time difference between the main pulse and afterpulse for the four channels. Right: Example of the HBT peak fit with a Lorentzian function with decay time of 135 ps convoluted with experimental resolution of 61 ps.}
\label{fig:HBT}
\end{center}
\end{figure}

We employed a commercial multi-channel infrared-sensitive SNSPD system (SingleQuantum EOS) and paired it with the quTAG TDC module with 7~ps timing resolution. The first step in the data processing was accounting for afterpulses, which likely appear because of non-ideal impedance matching in the front-end electronics. This lead to additional pulses delayed with respect to the primary signal by about 30~ns. Approximately 30\% of pulses were followed by an afterpulse, see the delay distributions in the left part of Figure~\ref{fig:HBT} for the precise fraction of afterpulses in each channel. To account for the afterpulses, all pulses registered within 35 ns of the previous hit in that channel were removed from the data.

To measure effects associated with change of the phase between the two thermal sources, which would model their relative displacement in space, we allowed the relative phase between two arms to evolve over duration of the experiment covering all possible values. The stability of the phase over the course of the approximately hour-long experiment was about 30 seconds. We therefore used correlations of coincidences between different channel pairs to study the effect.

As shown in Equation \ref{eq:outcomes_source_12} and explicated in \cite{stankus2020}, due to the two-photon amplitude interference effect, the probability of a coincident detection between two channels at opposite detection sites (L and R) is sensitive to the phase difference between photons arriving at each detection site. Moreover, when a phase difference causes an enhancement in channel pairs (cg) and (dh), it simultaneously causes a reduction in the number of coincidences in channel pairs (ch) and (dg), and vice versa. Therefore, the number of coincidences (in the HBT peak region) in channel pairs (cg) and (dh) are expected to be correlated with each other and anticorrelated with those in channel pairs (ch) and (dg). 

To test for these correlations between channel pairs, the number of coincidences within a $\pm 1.5 \sigma$ window of the central peak in the coincidence distribution (determined via a Gaussian fit of the distribution) was recorded over the duration of the hour-long experiment. This region is represented by the blue vertical bars in the right plot of Figure \ref{fig:HBT}. The number of coincidences within the central peak region was integrated over 20 seconds (roughly on the order of or smaller than the time scale of current phase stability) for each channel pair. 

\begin{figure}[h]
\begin{center}
\includegraphics[width=0.30\linewidth]{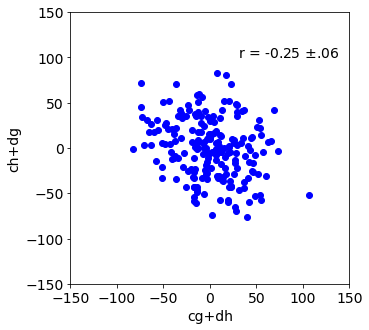}
\includegraphics[width=0.35\linewidth]{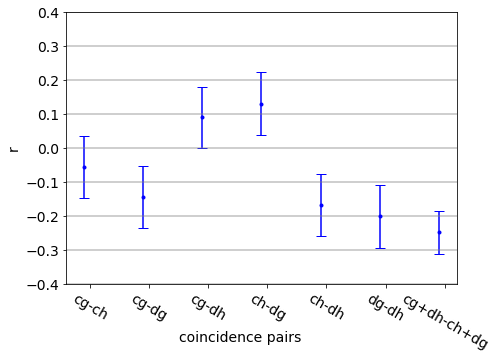}
\caption{Left: Scatter plot comparing the deviation from the mean number of coincident detections for channel pairs (cg) and (dh) versus (ch) and (dg). Resulting Pearson correlation coefficient is equal to $r = - 0.25 \pm 0.06$. Right: Correlations of coincidence rates for all possible combinations with corresponding uncertainties.}
\label{fig:correlations}
\end{center}
\end{figure}

Figure \ref{fig:correlations} shows a scatter plot comparing the deviation from the mean number of coincident detections for channel pairs (cg) and (dh) versus (ch) and (dg) to determine whether the expected anticorrelation is present. The Pearson correlation coefficient for the number of coincidences between these channel pairs was found to be $r = - 0.25 \pm 0.06$. Similar correlation analysis was conducted for each combination of channel pairs and the resultant correlations with corresponding uncertainties are displayed in the right image of Figure \ref{fig:correlations}. The correlations consistently show the correct signature as predicted by equation \ref{eq:outcomes_source_12} and were more than three standard deviations from a null result. 

Although in an ideal scenario an $r=-1$ is expected for anticorrelated channel pairs, there are factors in the experiment, such as shot noise and backgrounds, that dilute the observed correlation. Furthermore, since the phase variation is spontaneous, the phase may change from value to value within an integration window, further diluting the resultant anticorrelation. 

To test whether these are legitimate correlations that can be attributed to the two-photon amplitude interference effect and rule out systematic experimental bias, a similar analysis was performed on a region well outside the peak of the coincidence distribution, which is not expected to be sensitive to the phase difference. This region is indicated by the red bars in the right plot of Figure~\ref{fig:HBT}. Overall, analysis of coincidences in this region produced results consistent with the expected null-correlation, strengthening the conclusion that the correlations can be attributed to the two-photon amplitude interference effect. 

Another possible implementation of the experiment would require a deterministic variation of the photon phase difference so variations of the coincidence rates can be measured directly, as function of the phase. Work is in progress to stabilize the phase for sufficient time periods and to implement this scheme.

In conclusion, we described a novel approach to astrometry employing two stars as sources of single mode photons, which interfere and produce specific patterns of coincidences. We also presented  results from the first experiments as important step in establishing the HBT effect and its variation as function of the phase difference in the context of timing performance of the single photon detectors. Proof-of-concept experiments with two thermal sources and a variable phase delay are ongoing to demonstrate the predicted phase-dependent two-photon interference effects, eventually paving the way for testing this approach with sky sources.


\ack
This work was supported by the U.S. Department of Energy QuantISED award and BNL LDRD grant 19-30. A.P. and J.S. acknowledge support under the Science Undergraduate Laboratory Internships (SULI) Program by the U.S. Department of Energy. We thank Ryan Mahon for developing a fitting procedure. 
 
\section*{References}

\bibliographystyle{unsrt} 
\bibliography{report} 

\end{document}